\documentclass[useAMS,usenatbib]{mn2e}
\usepackage{epsfig}

\newcommand{\lsim}{\lower.5ex\hbox{$\; \buildrel < \over \sim \;$}}
\newcommand{\gsim}{\lower.5ex\hbox{$\; \buildrel > \over \sim \;$}}
\title[Annihilation of Positrons on PAHs in the ISM]{Positron Annihilation on Polycyclic Aromatic Hydrocarbon molecules in the Interstellar Medium}
\author[N. Guessoum, P. Jean, and W. Gillard]{N. 
Guessoum$^{1}$\thanks{E-mail: nguessoum@aus.edu, jean@cesr.fr, gillard@particle.kth.se}, P. Jean$^{2}$, and W. Gillard$^{3}$\\
$^{1}$American University of Sharjah, Physics Department, PO Box 26666, Sharjah, UAE\\
$^{2}$CESR, CNRS/Universit\'e de Toulouse, B.P. 4346, 31028 Toulouse Cedex 4, France\\
$^{3}$KTH, Department of Physics, AlbaNova University Centre, SE-10691 Stockholm, Sweden}

\begin{document}

\date{Accepted 2009 October 27. Received 2009 October 27; in original form 
2009 September 6}

\pagerange{\pageref{firstpage}--\pageref{lastpage}} \pubyear{2010}

\maketitle

\label{firstpage}

\begin{abstract}
   We examine the annihilation of positrons on polycyclic aromatic 
    hydrocarbon (PAH) molecules in interstellar medium conditions. 
    We estimate the annihilation rates of positrons on PAHs by a 
    semi-empirical approach. We show that PAHs can play a significant 
    role in the overall galactic positron annihilation picture and use 
    the annihilation rates and INTEGRAL galactic emission measurements 
    to constrain the amount of PAHs present in the ISM. We find an upper 
    limit of $4.6 \times 10^{-7}$ for the PAH abundance (by number, 
    relative to hydrogen).
\end{abstract}

\begin{keywords}
ISM: abundances -- ISM: molecules -- gamma-rays: theory.
\end{keywords}

\section{Introduction: Positron ISM Annihilation Issues and Interrogations}

In Astrophysics, the origin of positrons annihilating in the 
interstellar medium (ISM) and the nature and distribution of PAH 
molecules in space represent two unsolved problems which do not, at 
least on the surface, appear to have anything to do with one another. 

The abundant presence of positrons in our galaxy, especially in its 
bulge, and the high rate of annihilation determined through its 
gamma-ray line signature at 511 keV have been established for a few 
decades now \citep{Lev78, Pur97, Har98, Mil00, J03, K05}, after the 
initial observation of the radiation by \citet*{JHH72}, 
\citet*{JH73}, and \citet*{Hay75}. The flux and profile of the 
annihilation line, and thus the rate of steady production and 
annihilation of positrons have now been measured rather precisely by 
several instruments, balloon-borne and satellite-carried detectors 
(see, most recently, \citealt{Har98, K05, Wei08}). The analysis of 
the data on the gamma-ray emission (line and continuum) resulting 
from the annihilation of positrons on gas in various ISM regions has 
allowed for a better understanding of the processes and conditions of 
annihilation \citep{Chu05, J06}. Still, the origin of positrons has 
remained mostly a mystery, for the simple reason that no single 
astronomical population of objects is known to have both the spatial 
distribution that the annihilation radiation map presents and the 
capability to produce such high rates of positrons (for a brief 
review of the problem, see \citealt{GJP06}).

The story of the PAH molecules and their role in the ISM parallels 
that of the positrons. A few decades ago, the radiation emission 
signatures of these molecules, namely lines in the Infra-Red spectrum 
of dark nebulae, seemed to point to their abundant presence in the 
Galaxy (\citealt{Gil73} and others later). A large effort by 
researchers \citep{Dul81, Leg84, All85} and others later) helped 
identify characteristics of these molecules, including their 
structure, energy levels, and charge states; this was supposed to 
help pinpoint the presence and distribution of specific PAH molecules 
in different ISM regions. The latter goal has yet to be 
reached (see for instance \citealt{Rui05, Cam05}), so that even 
though it is now widely believed that roughly 10 \% -- perhaps more 
-- of the carbon in the ISM can be found in PAH molecules 
\citep{Tie90}, definitely identifying any of them has remained an 
elusive goal.

The connection between positrons and PAH molecules in earthly 
laboratories was made about a decade ago when experimental positron 
groups started to measure positron-molecule annihilation cross 
sections and realized that these increase by several orders 
of magnitudes when molecules are large \citep{Sur88, Iwa95, Iwa96, 
Bar03}. A substantial amount of experimental and theoretical work has 
been performed on this topic, and measurements of cross sections for 
a small dozen PAH and alkane molecules bombarded by positrons have 
been published, leading to theoretical efforts to explain the strong 
``Feshbach vibrational" resonance displayed in such interactions (see 
the extensive review of \citealt{Sur05}). A few researchers hinted 
at the possible importance of such reactions in the ISM, but no 
serious consideration of this issue has been attempted heretofore. 

The closest that astrophysicists came to addressing this issue was 
when the interaction of positrons with dust in the ISM was treated in 
detail \citep{GJG05}. In that work, dust was considered in the three 
forms it is usually presented in: ``big grains", ``very small 
grains", and PAHs. And because the positron--dust-grain cross section 
was assumed to be essentially geometric, ``very small grains" were 
neglected next to ``big grains", which themselves proved to be 
relevant only in the hot phases of the ISM and, in some special 
conditions, in the warm ionized phase. It seemed reasonable then to 
conclude that PAHs, being even smaller and much less abundant, would 
play a negligible role in the positron's life and death in the ISM. 
But that was not correct, since the positron--PAH cross section was 
not geometrical but rather highly resonant, especially at very low 
energies/temperatures; indeed, with cross sections sometimes a 
million times larger than those of charge exchange with hydrogen 
atoms, abundances of PAHs of about $10^{-6}$ (by number, N$_{\rm 
PAH}/$N$_{\rm H}$) would make them roughly as important as other 
species. 

We must stress, however, that PAHs, although commonly considered as 
the ``molecular end" of the dust grain size distribution, behave very 
differently with respect to positrons. That is why the treatment of 
positron annihilation on PAHs is warranted now and is fundamentally 
distinct from all past (astrophysical) works.

In this work, we investigate the extent to which PAH molecules are 
relevant in the positrons' annihilation in the ISM. We have thus 
assembled the relevant information regarding PAHs in the ISM 
(abundances, sizes, distributions, charge states, etc.), and on the 
interactions between positrons and PAHs. With that we estimate the 
rate of positron annihilation on PAH molecules in the various ISM 
phases and compare it to those of other processes \citep{GJG05}.

In the next section we briefly review current knowledge about 
PAH molecules in the ISM. In section 3 we present the main 
experimental information on positron annihilation on PAHs and then 
proceed to determine the positron--PAH reaction rate at 8000 K  
(i.e. in the warm media) by several steps. In section 4 we use that 
knowledge to calculate the rate of annihilation of 
positrons on PAHs in the relevant phases of the ISM, paying 
particular attention to the PAH charging effect. In section 5 we 
explore the observational consequences of our 
calculations using the measurements performed with the spectrometre 
SPI 
on the space observatory INTEGRAL (INTErnational Gamma-Ray 
Astrophysics Laboratory).
In section 6 we summarize our findings and present our main 
conclusions 
regarding the possible importance of this process in current positron 
astrophysical studies and point to the work that will be needed in 
the 
future in order to advance our 
understanding of this problem.

%%%%%%%%%%%%%%%%%%%%%%%%%%%%%%%%%%%%%%%%%%%%%%%%%%%%%%%%%%%%%%%%%%%%%%%%%%
% 2. PAHs in Astrophysics
%%%%%%%%%%%%%%%%%%%%%%%%%%%%%%%%%%%%%%%%%%%%%%%%%%%%%%%%%%%%%%%%%%%%%%%%%%

\section[]{PAHs in Astrophysics}

Polycyclic Aromatic Hydrocarbons are organic molecules which 
consist entirely of C and H atoms and have a polycyclic structure, 
which 
makes the energy needed for breaking them up much higher than 
average; this then helps them survive in space. In addition, each 
ring has 6 electrons from the carbon atoms ``floating" around and 
contributing to the binding. These molecules have ``aromatic" 
properties. 
The prototype example of a PAH is Naphthalene ($\rm C_{10}H_8$), 
which is made of two rings. A simpler example would be benzene ($\rm 
C_6H_6$), but it consists of only one ring, so it is not 
``polycyclic". 

These molecules became the focus of important astrophysical research 
when the ``unidentified infrared (UIR) emission bands" (3-13 $\mu m$) 
that had been observed from nebulae since the early seventies were 
shown \citep{Leg84, All85}, more than a decade later, to be very 
similar to those produced by PAHs when they are temporarily heated 
(in space by UV radiation); UIR bands then came to be referred to as 
AIBs (aromatic IR bands). Indeed, the usual dust grain populations 
could not be responsible for such emissions because they were too big 
and thus could not be heated to 1000 K or more and cooled quickly. 
Using observations made by space-borne IR telescopes, similar IR 
emission 
bands were later found in a variety of objects, ranging from comets 
to galaxies (e.g. \citealt{Ehr02}). \citet*{Sal99} concluded that 
``PAHs are ubiquitous throughout the general diffuse ISM". Such 
identifications allowed for an inference of the abundance of PAHs in 
the ISM, roughly believed to be $~ 10^{-7}-10^{-6}$ (by number, i.e. 
N$_{\rm PAH}/$N$_{\rm H}$) \citep{All96}, making them the most 
abundant molecules in the ISM 
after H$_2$ and CO.

PAHs are often seen as an extension of, or even the seeds for, the 
small dust grain populations \citep{All96, Sal99, Pee02, Abe05}. 
Indeed, not only are they the 
largest molecules known in space, but the largest of them have been 
described \citep{Pee02, Rap05} as aggregates of planar 
molecules, which when stacked can become the ``very small grains" of 
dust, which range in size from 1 nm to 10 nm. In fact, it is 
postulated that PAHs constitute the first step in the formation of 
dust grains. 

The above-mentioned emission lines are generally associated with C-C 
and 
C-H vibrational modes of PAHs of various sizes (\citealt{Dul81} 
first, and many researchers subsequently). It is also often concluded 
that PAHs with a rather large number of carbons (greater than about 
30) are responsible for some of the AIB features, particularly those 
around 6.2 $\mu$m and 11.3 $\mu$m. Since the photodissociation 
threshold energy is a few tenths of eV per carbon atom, it is clear 
that ``small" molecules (with a number of carbon atoms N$_C \lsim 
30$) would more easily be 
depleted in H II regions by energetic photons \citep{All96, Pee02, 
Abe05}; bigger ones easily survive. It 
has also been postulated that UV, cosmic rays, and shock waves could 
chemically alter such molecules, particular in warm/hot, 
collisionally active environments. Finally, PAHs can be neutral or 
electrically charged, depending on the densities of the photoionizing 
radiation and of free electrons \citep{Omo86, Dra87, Lep88, Bak94, 
All96, Dar97, Wei01a, Wei01b, Abe05}, and their IR emission spectra 
are then significantly different.

%%%%%%%%%%%%%%%%%%%%%%%%%%%%%%%%%%%%%%%%%%%%%%%%%%%%%%%%%%%%%%%%%%%%%%%%%%
% 3. Laboratory Data on e+ annihilation on PAHs 
%%%%%%%%%%%%%%%%%%%%%%%%%%%%%%%%%%%%%%%%%%%%%%%%%%%%%%%%%%%%%%%%%%%%%%%%%%

\section{Current Laboratory Data on Positron Annihilation on PAHs}

As mentioned in the introduction, interest in e$^+$--PAH annihilation 
grew with the development of experimental techniques that allowed for 
the
measurement of cross sections of positrons annihilating with more and 
more complex atoms and molecules \citep{Iwa96, Iwa97, Iwa00, Sur00, 
Gil00, Sul02, Bar03, Mar04, Sur05, Sur06}. Indeed, it soon became 
apparent that  
e$^+$--PAH reactions exhibited huge cross 
sections, or equivalently reaction rates or $Z_{eff}$, the latter 
being defined as:
\begin{equation}
 Z_{eff} = {\lambda \over {\pi r_0^2 c}} \; ,
\end{equation}
\noindent where $\lambda$ is the annihilation rate (in cm$^3$/s), and 
$r_0$ is the classical electron radius; it turns out that $Z_{eff}$ 
has essentially nothing to do with the number of electrons Z of the 
species (for example, $Z_{eff}$ = 8 for H), and indeed reaches values 
of $10^7$ or more for PAH molecules (see the references given above). 
However, only relatively small PAH molecules have been experimentally 
investigated so far \citep{Iwa96, Iwa97, Bar03}, as molecules with 
more than 3 aromatic rings have such low vapor pressures that 
performing those measurements is a great experimental challenge. The 
only PAHs for which a $Z_{eff}$ was actually experimentally measured 
\citep{Iwa96, Iwa97} are listed in Table 1. In the \citet*{Iwa96} 
experiment the FWHM of the annihilation line resulting from positrons 
on Naphthalene was also measured and found to be 2.29 keV (compared 
to 1.71 keV for H$_2$). 

%%%%%%%%% Table 1: Measured Z_eff for PAH molecules %%%%%%%%%%%%%%%%

\begin{table}
    \caption{Measured $Z_{eff}$ for PAH molecules at room temperature 
(Iwata et al. 1996; Iwata et al. 1997).}
    \begin{array}[b]{llc}
	\noalign{\smallskip}
	\hline
	\noalign{\smallskip}
      \mbox{Molecule} & \mbox{Chemical Formula} & \mbox{$Z_{eff}$} \\
	\noalign{\smallskip}
	\hline
      \hline
	\noalign{\smallskip}
      \mbox{Benzene}  & \mbox{C$_{6}$H$_6$} & \mbox{18000} \\
      \mbox{Toluene}   & \mbox{C$_{7}$H$_{8}$} & \mbox{190000} \\
      \mbox{Naphtalene} & \mbox{C$_{10}$H$_{8}$} & \mbox{494000} \\
      \mbox{Anthracene} & \mbox{C$_{14}$H$_{10}$} & \mbox{4330000} \\
	\noalign{\smallskip}
     \hline \\
      \end{array}
\end{table}

%%%%%%%%%%%%%%%%%%%%%%%%%%%%%%%%%%%%%%%%%%%

The very large values for $Z_{eff}$ have been theoretically 
interpreted in terms of resonant interaction between the positron and 
the molecule, or equivalently by the formation of temporary 
positron-molecule bound states. \citet*{Gri00} has come up with a 
theoretical model to reproduce the experimental data on the cross 
sectional profiles in terms of the positron momentum distribution and 
of the temperature of the medium; he also produced a simple fit for 
the exponential increase of $Z_{eff}$ in terms of the number of atoms 
in the aromatic molecules (he found $Z_{eff} \propto N^{8.2}$, where 
N is the total number of atoms in the molecule). We note, however, 
that $Z_{eff}/Z$ tends to saturate for larger N's in other types of 
molecules (see, for instance, the data on alkanes in Fig. 8 of 
\citet*{Bar03}). None of the studies which investigated the 
variations of 
$Z_{eff}$ with the physical characteristics of molecules could 
satisfactorily reproduce the saturation effect for large molecules 
\citep{Mur91, Lar97, Lar98, Iwa00}.
This saturation is clearer when values of 
$Z_{eff}$/Z are presented as a function of the number of electrons Z 
of 
the molecules (see \citealt{Iwa95} and Fig. 39 of \citealt{Sur05}) 
and we thus propose a different fit. We have found the 
relation :
\begin{equation}
ln(\frac{Z_{\rm eff}}{Z}) = A\left( 1 - e^{-\frac{Z}{B}} \right)
\end{equation}
\noindent to give a satisfactory fit (see Fig. 1); the number of 
measurements is much greater for alkanes than for PAHs; we show that 
$Z_{eff}$'s for the two types of molecules converge in this 
representation. In our fitting function (Eq. 2), the best parameter 
values
are: $A = 10.75 \pm 0.44$, $B = 40.1 \pm 3.8$. 
We hasten to add that this relation was, like Gribakin's, not based 
on  
any physical considerations, only the fact that experiments show that 
$Z_{eff}/Z$ levels off for large molecules. We should add that 
e$^-$--PAH cross sections, being non-resonant, are not of much help 
in this regard; 
like Gribakin, our only recourse was an empirical fit.

%%%%%%%% Figure 1: ln(Z_eff/Z) for Alkanes and PAHs %%%%%%%%
\begin{figure*}
\centering
\includegraphics[width=11.73cm,height=8.0cm]{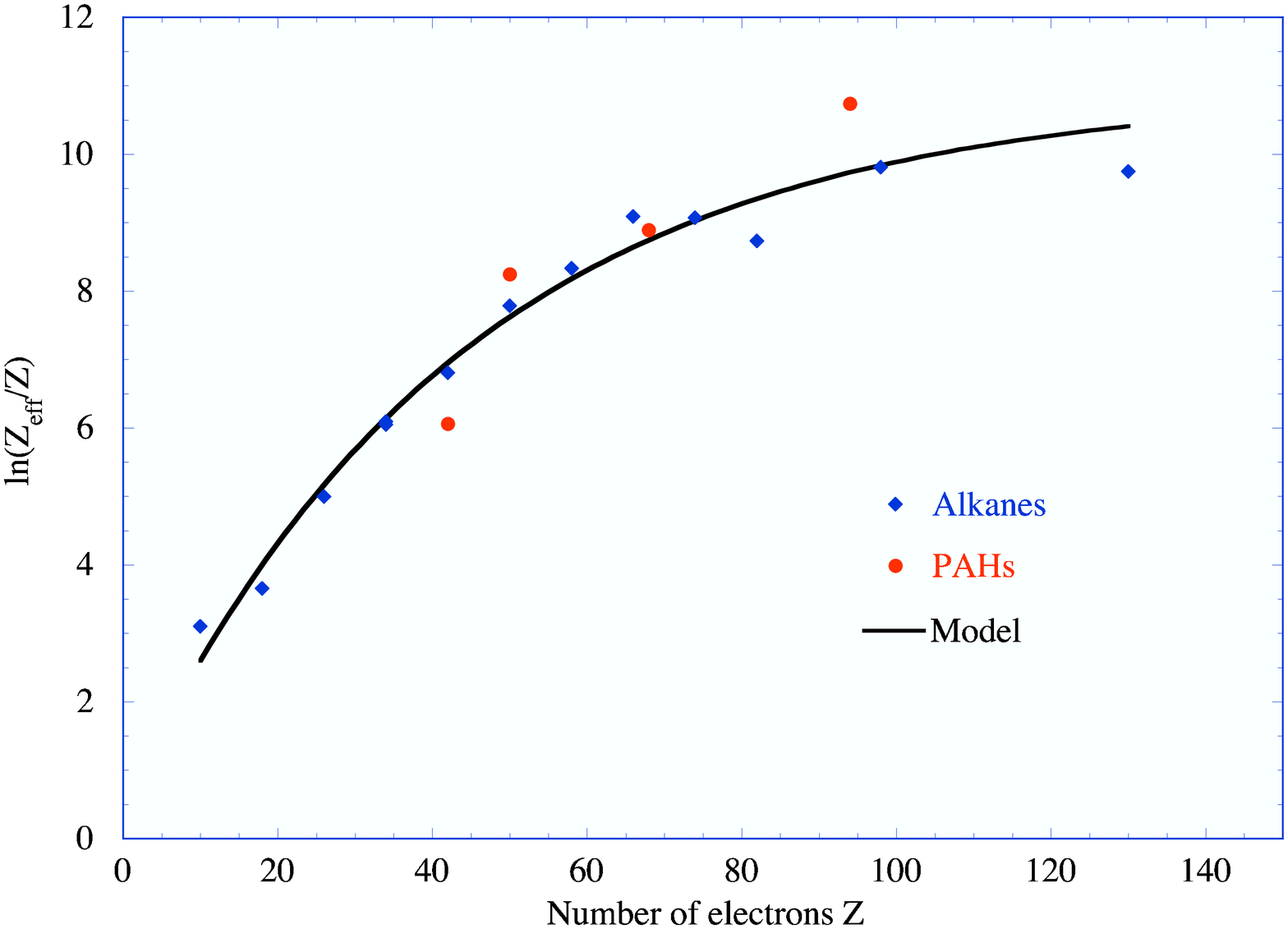}
\caption{Comparison of ln($Z_{eff}$/Z) for alkanes and PAHs as 
functions of the number of electrons in the molecule Z (Iwata et al. 
1995). Also shown is the best fit of the empirical function (see Eq. 
1) obtained with $A$ = 10.75 and $B$ = 40.1.}
\end{figure*}
%%%%%%%%%%%%%%%%%%%%%%%%%%%%%%%%%%%%%%%%%%%

We must emphasize that the experimental values of $Z_{eff}$ were 
obtained at room 
temperature only, whereas the temperature we are mainly interested 
in is 8000 K, since positron annihilation in the ISM is believed to 
occur 
mostly in the warm (neutral and ionized) phases, where T $\approx 
8000$ K 
\citep{Chu05, J06} and PAHs are expected to be easily evaporated in 
the hot phase (where T $\sim 10^6$ K). To our knowledge, only one 
experiment has measured $Z_{eff}$ 
as a function of T \citep{Iwa00}, but this was limited to very 
few and small molecules (methane CH$_4$, ethylene C$_2$H$_4$, and 
butane C$_4$H$_{10}$) and to temperatures up to about 2500 K. More 
importantly, we note that measurement errors are largely unknown 
(rarely, if ever, given in the scant literature); consequently, the 
uncertainties we have given above (for $A$ and $B$) are related 
to the fit only, and the few measurements that are available do not 
allow us to constrain the ``saturation" of the $Z_{eff}$ function (in 
terms of Z) very well. We do insist, however, on the existence of 
such a saturation, unlike what the Gribakin function implies, and 
that in turn has important consequences on the relative role of 
PAH's in the annihilation of positrons in the ISM.

So we need to determine the value of $Z_{eff}$ at 8000 K, for which 
we need measurements of the cross section as a function of the 
positron energy. Some experiments (see \citealt{Bar03} and 
\citealt{Sur05}) have performed such measurements, so we can simply 
calculate $Z_{eff}$ as a function of the positron temperature by way 
of the reaction rate $\lambda$:
\begin{equation}
\lambda = \langle \sigma v\rangle = \int_{0}^{\infty}{2 \over 
\sqrt{\pi}} 
{{\sqrt{E}\over {(kT)}^{3/2}} e^{-E/kT} \sigma(E) v dE} \; , 
\end{equation}
\noindent where $\sigma$ is the annihilation cross section, v the 
positron velocity, and E its kinetic energy.

Noting that at low temperatures $Z_{eff}$ drops as $T^{-1/2}$ (see 
Figure 8 of \citealt{Iwa00}), the above approach would yield the 
right behavior for $Z_{eff}$ only if the cross sections were extended 
at very low energies as $E^{-1}$. We were thus able to reproduce the 
experimental data of $Z_{eff}(T)$ for butane rather well for most 
temperature ranges, although $Z_{eff}$ tends to decline very slowly 
at higher temperatures and not flatten out completely. Figures 2 (a 
and b) show the behavior of $Z_{eff}$   as a function of temperature 
for various molecules; Figure 2b shows the variation of normalized 
values of $Z_{eff}$ on a linear scale so as to emphasize the weak 
dependence of $Z_{eff}$ on $T$.

Having noted the variation of $Z_{eff}$ with temperature for alkanes, 
which we assume to be the same for PAHs, we then produce values of 
$Z_{eff}$ at 8000 K for a few PAHs using Figure 1 and the 
T-dependence. In Table 2 we give values of $Z_{eff}$ for some alkane 
and PAH molecules.

%%%%%%%%% Table 2: Calculated Z_eff for a few alkanes and PAH molecules 
%%%%%%%%%%%%%%%%

\begin{table}
    \caption{Calculated $Z_{eff}$ at T = 8000 K for a few 
(medium-large) Alkane (top group) and PAH molecules (bottom group).}
    \begin{array}[b]{llc}
	\noalign{\smallskip}
	\hline
	\noalign{\smallskip}
      \mbox{Molecule} & \mbox{Chemical Formula}      &  
\mbox{$Z_{eff}$}    \\
	\noalign{\smallskip}
	\hline
      \hline
	\noalign{\smallskip}
      \mbox{Butane}  	  &   \mbox{C$_{4}$H$_{10}$}	 & \mbox{ $6.4\times10^3$}	 \\
      \mbox{Hexane}  	  &   \mbox{C$_{6}$H$_{14}$}	 & \mbox{$5.5\times10^4$ }	 \\
      \mbox{Heptane}      &   \mbox{C$_{7}$H$_{16}$}    & \mbox{$1.8\times10^5$ }	 \\
      \mbox{Octane}       &   \mbox{C$_{8}$H$_{18}$}    & \mbox{ $3.5\times10^5$}	 \\
      \mbox{Nonane}       &   \mbox{C$_{9}$H$_{20}$}    & \mbox{ $6.8\times10^5$}	 \\
      \mbox{n-Dodecane}   &   \mbox{C$_{12}$H$_{26}$}    & \mbox{$2.6\times10^6$ }	 \\
	\noalign{\smallskip}
	\hline
	\noalign{\smallskip}
      \mbox{Naphthalene}  &   \mbox{C$_{10}$H$_8$}	 & \mbox{4.2 $\times$ 10$^5$}	 \\
      \mbox{Anthracene}   &   \mbox{C$_{14}$H$_{10}$}	 & \mbox{4.1 $\times$ 10$^6$}	 \\
      \mbox{Hexacene}     &   \mbox{C$_{26}$H$_{16}$}    & \mbox{2.4 $\times$ 10$^7$}	 \\
      \mbox{Octacene}     &   \mbox{C$_{34}$H$_{20}$}    & \mbox{1.1 $\times$ 10$^8$}	 \\
      \mbox{Decacene}     &   \mbox{C$_{42}$H$_{24}$}    & \mbox{1.5 $\times$ 10$^8$}	 \\
 	\noalign{\smallskip}
     \hline \\
      \end{array}
\end{table}

%%%%%%%%%%%%%%%%%%%%%%%%%%%%%%%%%%%%%%%%%%%

We emphasize the large uncertainties inherent in these calculations 
due to the fact that only a few actual measurements exist for 
positron annihilation reactions with alkanes and/or small PAHs at low 
temperatures. We call for new experiments and extensive 
measurements as well as theoretical calculations of annihilation 
cross sections for positrons with as many PAH species as possible.

%%%% Figure 2: Z_eff's for Alkanes and PAHs as a function of temperature %
\begin{figure*}
\centering
\includegraphics[width=11.7cm,height=8.0cm]{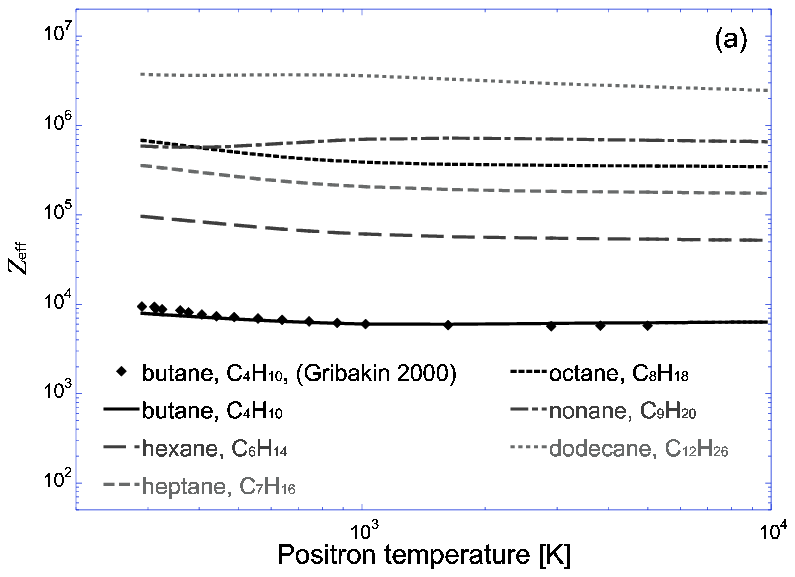}
\includegraphics[width=12.5cm,height=8.0cm]{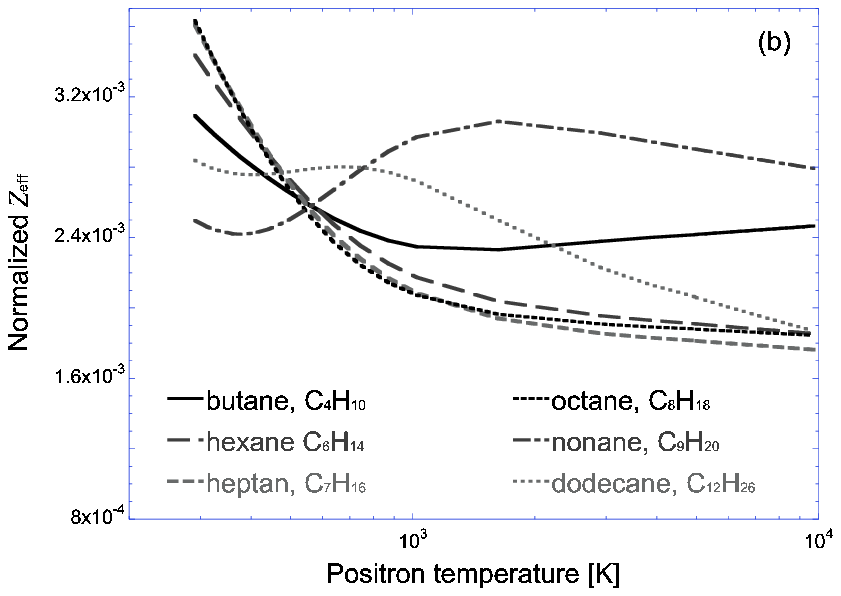}
\caption{$Z_{eff}$ as a function of temperature for various molecules 
(alkanes and PAHs) in (a) logarithmic scale and (b) normalized to the 
value 1.}
\end{figure*}
%%%%%%%%%%%%%%%%%%%%%%%%%%%%%%%%%%%%%%%%%%%

\citet*{Bar03} measured the rate of annihilation of 
low energy positrons with many hydrocarbon molecules. They found that 
$Z_{eff}$ is enhanced at positron energies that correspond to 
resonance energies of the strongest infrared active vibrational 
modes. They interpreted these results as excitations of vibrational 
Feshbach resonances by incoming positrons which are temporarily 
trapped on the molecules. Based on this observation, it would in 
principle 
be possible to infer the variation of the annihilation rate on PAHs 
as 
a function of the temperature of thermalized positrons by integrating 
the product of their (Maxwellian) velocity distribution with the 
spectrum 
of the vibrational modes of the PAH molecules. 
However, the observed resonance effect does not seem to 
be systematic. Indeed, there is no clear enhancement of $Z_{eff}$ for 
some energies of positrons (e.g. see Figs. 12 and 13 of 
\citealt{Bar03}). Moreover, a shift in the position of the resonance 
with respect to the vibrational mode is observed, which 
seems to be linked to the binding energy of the positron to the 
molecule.

Referring to recent work on energy-resolved positron-molecule 
annihilation cross sections \citep{Bar06}, Surko 
has remarked (private communication) that the temperature dependence 
of $Z_{eff}$ referred to above ($Z_{eff} \propto T^{-1/2}$ obtained 
by \citealt{Iwa00} up to 2500 K) may not hold at 8000 K, since the 
$1/E$ variation of the cross section, which leads to the $T^{-1/2}$ 
dependence of $Z_{eff}$, seems to apply only at very low energies ($E 
<< 
E_{res}$), that is for direct annihilation, whereas for ``high" 
temperatures, resonant effects change that behavior. Should $Z_{eff}$ 
decrease faster with $T$, the values at 8000 K would be significantly 
smaller than inferred here. However, this uncertain behavior is 
itself unclear, as we do not know the binding energies for the larger 
PAHs and thus the energy difference between those and the vibrational 
energy values. This shows that additional (unknown) factors need to 
be taken into account in order to build a correct model. 

Consequently, we cannot yet analytically obtain $Z_{eff}$ as a 
function of the temperature and of N (the number of atoms) in large 
PAHs with our current level of understanding of the annihilation of 
positrons with molecules. We are then left with the necessity of 
producing a numerical fit of the available data for $Z_{eff}$(N) and 
then extrapolating for larger molecules, since it turns out that the 
prime contributors for positron annihilation on PAHs in the ISM are 
those with a number of atoms around 50. (See below the 
discussion about PAH size distributions and the contribution of each 
species to the annihilation rate.)

Since the larger PAH molecules will be the more relevant ones for our 
astrophysical positron considerations, we believe that this issue of 
$Z_{eff}$ for PAHs with $N_C \gsim 30$ ($N \gsim 50$) is an important 
one that should be attempted experimentally at the earliest.

%%%%%%%%%%%%%%%%%%%%%%%%%%%%%%%%%%%%%%%%%%%%%%%%%%%%%%%%%%%%%%%%%%%%%%%%%%
% 4. Annihilation of Galactic Positrons on PAHs (relative galactic importance)
%%%%%%%%%%%%%%%%%%%%%%%%%%%%%%%%%%%%%%%%%%%%%%%%%%%%%%%%%%%%%%%%%%%%%%%%%

\section{Annihilation of Galactic Positrons on PAHs (relative 
galactic importance)}

We now set out to determine the rate of positron annihilation on PAH 
molecules in the ISM and compare it with those of positrons with the 
usual gas components 
(free electrons, atomic and molecular hydrogen, helium, dust grains). 
We may start from the reaction rate formula $\lambda = Z_{eff} \pi 
r_0^2 c$ defining $Z_{eff}$, and sum up the contributions of the 
various PAH species through their number-abundances $y_s$ (the 
subscript s 
referring to any given species), considering that some may be 
partially destroyed, depending on the physical conditions in the 
medium. One must also take into account the possible electric charge 
of the molecule, which would then enhance or reduce the ``affinity" 
between the positrons and the molecules. 

The reaction rate for each PAH species can then be written as:
\begin{equation}
 \lambda_{e^+-s} = y_s Z_{eff,s} \pi r_0^2 c f_{elec-s} \; ,
\end{equation}
\noindent and this would then be summed over for the various PAH 
species. 
Note that $y_s$ includes the destruction effects; that is, $y_s = 
y_{s0} 
\times f_{dest-s}$, where $y_{s0}$ represents the abundance of an 
s-species of PAH in the absence of any destruction effects; 
$f_{dest-s}$ 
and $f_{elec-s}$ will be estimated in the next sub-section.

\subsection{PAH charging and destruction:}

Several studies have been performed on the determination of the 
charge states of dust grains and PAH molecules of various sizes in 
different conditions and by several physical processes. 
\citet*{Dra87} considered collisional effects and showed that charge 
states other than 0, +1, and -1 have extremely low probabilities 
($\sim 
10^{-4}$ or less) in various physical conditions. This confirmed 
statements by \citet*{Omo86}. \citet*{Lep88} considered diffuse ISM 
clouds and determined charge states in various regions of the cloud 
(edge, 
center, etc.); they concluded that the neutral state is most 
prevalent.
\citet*{Bak94} then added 
photoelectric effects to the analysis and calculated the f(Z) 
probability of finding a grain/PAH molecule in a charge state Z. 
Using typical densities, temperatures, and UV field intensities in 
various phases of the ISM, they found (their figure 6): charge states 
of mostly 0 and 1 (average of $~ 0.5$) and -1 and 0 (average of 
$\approx 
-0.25$) for the Warm and Cold ISM media, respectively. The work of 
these 
authors was revisited by \citet*{Dar97}, who applied their 
ionization model to various conditions of the ISM and of nebulas; 
they 
found that in the diffuse ISM, 65 \% of large 
PAH molecules are negatively charged and 35 \% are neutral (different 
results are obtained for different physical conditions). 
Photoelectric 
charging was revisited by \citet*{Wei01a} with improved 
atomic data, grain size distributions, etc.; they found (their figure 
9) that except for very large PAH molecules in the warm medium, where 
the 
average charge was $\approx 0.4$, all PAH's, especially mid-size 
molecules, will be neutral in practically all conditions. 

As to how the charge state of a PAH molecule will affect the reaction 
rate of capture and annihilation of a positron, we refer to the 
thorough discussion of the physics and chemistry of interstellar PAHs 
by \citet*{Omo86}. This author tells us that because the charge of a 
PAH is 
essentially distributed over its entire surface, the processes which 
govern the charges of interstellar PAH's are very similar to those 
involved in the charging of interstellar dust grains. He thus 
concludes that one may reasonably adapt the classical discussions of 
the charging of interstellar grains to PAH's, basing this conclusion 
on several works (see references therein).

The positron-charged-PAH effect on the annihilation rate can then be 
dealt with in the same way as we did in \citet*{GJG05}, i.e. as did 
\citet*{Zur85} and \citet*{GRL91}. That is, the reaction rate gets 
multiplied by an ``electric charge factor", f$_{elec}$, where 
f$_{elec}$ is given by $(1 - Ze^2/a_skT)$ or $\exp(-Ze^2/a_skT)$ 
depending on whether the molecule, with a radius $a_s$, is negatively 
or positively charged, and where $f_{elec}$ then takes the values of 
3.1 and 0.12 in the Z = -1/+1 charge state, for a typical PAH size of 
1 nm. 

As to the PAH destruction effect, we simply consider that in the hot 
phase of the ISM, electrons are more than energetic enough ($kT$ 
$\sim 100$ eV) to break the molecule every time a collision occurs, 
so that the PAHs 
are largely depleted from such an environment; however, in the warm 
and cold phases ($T \approx 8000$ K and 10-100 K respectively), 
collisions will rarely, if ever, destroy the PAH molecules; therefore 
$f_{dest}$ = 0 for the hot phase and $f_{dest}$ = 1 for the others.

\subsection{e$^+$-PAH reaction rate in the ISM:}

Considering that the PAH number-abundances of $~ 10^{-7}-10^{-6}$ 
quoted 
above are for all PAH molecules in the ISM, and noting (from Figure 
1) that only large molecules will be effective annihilators of 
positrons, we must convolve the $Z_{eff}$ values shown in Figure 1 
with the PAH size or atom-number distribution; \citet*{Des90} and 
\citet*{DL98} give somewhat different PAH distributions, which 
\citet*{Pil09} present (their Figure 5) as a function of the number 
of carbon atoms in the molecule. The rates obtained using the two 
distributions differ by only 0.9\%; in the rest of this work, we 
adopt the more recent, ``lognormal" distribution of \citet*{DL98}.

Putting all factors together then gives the following expression for 
the reaction rate between positrons and PAHs:
\begin{equation}
      \lambda_{e^+-PAH} =  \left[ \int_N Z_{eff}(N) dy_s(N) \ 
f_{elec-s} \right] \pi c r_0^2 \ Y_{PAH}   \; ,
\end{equation}
\noindent where $Y_{PAH}$ refers to the total PAH abundance (relative 
to H) in the ISM. The summation over species then gives:
\begin{equation}
     \lambda_{e^+-PAH} \approx  1.5 \times 10^{-13} \ {Y_{PAH} \over 
10^{-6}} \ \langle f_{elec} \rangle \; {\rm cm}^3/{\rm s} \; .
\end{equation}

We should note, however, that taking into account the uncertainty on 
the 
semi-empirical model of $Z_{eff}$ (see section 3) the relative 
uncertainty on the derived rate is about 47\%. 

We now compare this result with the rates for the main positron 
processes in the various ISM phases (charge exchange with H, direct 
annihilation with free and bound electrons, radiative combination 
with electrons, capture by dust grains) as given in \citet*{GJG05}; 
the values for all rates are given in Table 3. 
We first note that if the total PAH abundance in the ISM is of the 
order of 10$^{-7}$ or less, then positrons will not be affected by 
PAH molecules, however large they may be. If, however, $Y_{PAH} \sim 
10^{-6}$, then the rate of annihilation of positrons by PAHs becomes 
non-negligible, especially if the molecules are neutral or negatively 
charged. Indeed, in the warm neutral phase this process becomes 
second in importance, with a rate up to 26 \% that of charge exchange 
with hydrogen when PAHs are negatively charged. Similarly, in the 
warm 
ionized phase the process is then second in importance, with a rate 
up to 39 \% that of radiation combination with free electrons. 
We must stress, however, that it is generally believed that 
PAHs would normally exist in multiple charge states, with perhaps a 
predominance of neutral molecules -- thus reducing their importance 
as 
annihilators.

%%%%%%%%% Table 3: Reaction rates for positron annihilation by various processes %%%%%%%%%%%%%%%%

\begin{table*}
\centering
\caption{Reaction rates for positron annihilation (by various 
processes) in the warm neutral (WNM) and ionized (WIM) phases of the 
ISM. The reaction rate for positron annihilation with PAHs is 
calculated using the total PAH number-abundance $Y_{PAH}=10^{-6}$}
    \begin{array}[t]{lcc}
	\noalign{\smallskip}
	\hline
	\noalign{\smallskip}
      \mbox{Process}	&  \mbox{$r_{e^+-s}$ (cm$^3$/s)}  &  
\mbox{$r_{e^+-s}$ (cm$^3$/s)}   \\
      			& \mbox{WNM}  & \mbox{WIM} \\
	\noalign{\smallskip}
	\hline
	\hline
	\noalign{\smallskip}
      \mbox{Charge exchange with H} 	& \mbox{1.8 $\times$ 10$^{-12}$}	
& \mbox{ -- } \\
      \mbox{Direct annihilation with free electrons}  & \mbox{	--	} & 
\mbox{1.7 $\times$ 10$^{-13}$}  \\
      \mbox{Direct annihilation with bound electrons} & \mbox{4.4 
$\times$ 10$^{-14}$} &	--	\\
      \mbox{Radiative combination with electrons}     & \mbox{	--	} & 
\mbox{1.2 $\times$ 10$^{-12}$ }  \\
      \mbox{Capture by dust grains}	& \mbox{6.5 $\times$ 10$^{-15}$} 
& \mbox{4.6 $\times$ 10$^{-14}$} \\
      \mbox{Annihilation with PAH molecules} 	& \mbox{1.5 $\times$ 
10$^{-13}$} & \mbox{1.5 $\times$ 10$^{-13}$}	 \\
	\noalign{\smallskip}
     \hline \\
      \end{array}
\end{table*}

%%%%%%%%%%%%%%%%%%%%%%%%%%%%%%%%%%%%%%%%%%%

%%%%%%%%%%%%%%%%%%%%%%%%%%%%%%%%%%%%%%%%%%%%%%%%%%%%%%%%%%%%%%%%%%%%%%%% 

% 5. Observational Consequences
%%%%%%%%%%%%%%%%%%%%%%%%%%%%%%%%%%%%%%%%%%%%%%%%%%%%%%%%%%%%%%%%%%%%%%%% 

\section{Observational Consequences}

One observational consequence of our calculation of the rate of 
positron annihilation on PAHs is the possible enhanced effect these 
molecules would have on the 511 keV line produced in a nebula where 
these molecules may be abundant, especially if physical conditions 
(e.g. high collision rates) lead to a significant negative charging 
of 
the PAHs. And should PAH molecules, in such situations, start to 
dominate 
in annihilating positrons, the observed 511 keV spectra 
would be different from the usual ``general ISM" ones. Indeed, since 
measurements of the spectra of the annihilation of positrons with PAH 
molecules \citep{Iwa97} have determined FWHM's of 
the annihilation line that are between 2.0 and 3.0 keV, the spectra 
from PAH-rich and negatively charged nebulae would be substantially 
wider than one recorded from a warm region of the ISM, where a line 
of width $~ 1.5$ keV is produced. Furthermore, the positronium 
fraction (or $3\gamma / 2 \gamma$ ratio) would likely be different in 
the two measurements. 

Using the method presented in \citet*{GJG05} we have calculated the 
spectral distribution of the annihilation emission including the 
contribution of PAHs with an abundance of $Y_{PAH} \sim 10^{-6}$, 
taking their electric charge into account. The FWHM of the 
annihilation line in PAH was here taken to be $\approx$ 2.5 keV.
In the warm neutral medium, the positronium fraction does not change 
significantly: from 99.9 \%\ in the absence of PAHs to 99.4 \%\ and 
98.6 
\%\ with neutral and negatively charged PAHs, respectively.
In the warm neutral medium, the shape of the 511 keV line is not 
notably modified. 
The situation is quite different in the warm ionized medium. The 
positronium fraction decreases from 87 \%\ without PAHs to 78~\%\ and 
64 
\%\ with neutral and negatively charged PAHs, respectively. The 
influence 
of PAHs distinctly deforms the base of the spectral line (see Fig. 4).

%
%%%%%%%% Figure 3: Model of spectral distribution WIM %%%%%%%%
\begin{figure*}
\centering
\includegraphics[width=11.73cm,height=8.0cm]{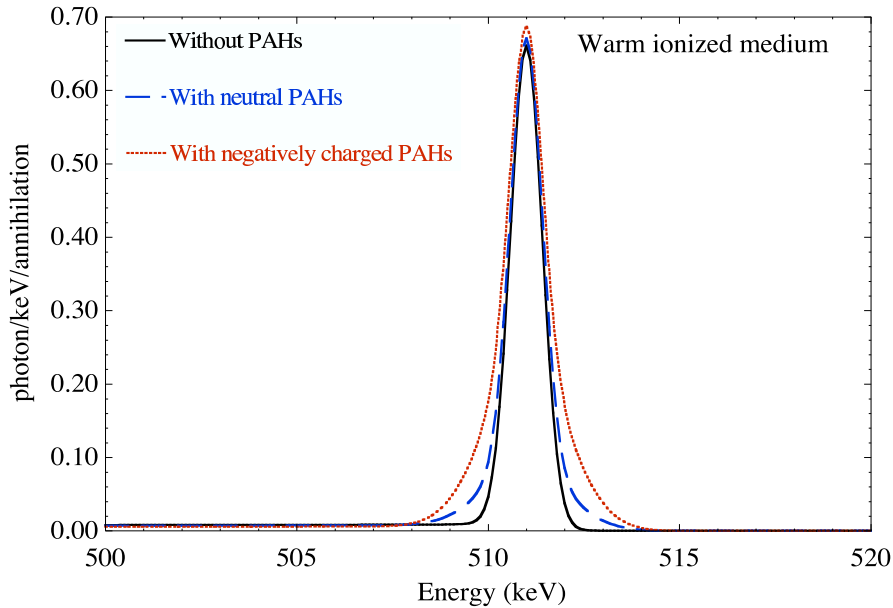}
\caption{Spectral distribution of the annihilation emission in the 
warm ionized medium with and without PAH contribution, calculated 
assuming an abundance of PAHs of $Y_{PAH} \sim 10^{-6}$. The cases of 
neutral and negatively charged PAHs are presented separately.}
\end{figure*}
%%%%%%%%%%%%%%%%%%%%%%%%%%%%%%%%%%%%%%%%%%%

Such a test cannot be undertaken at present 
because INTEGRAL-SPI does not have pinpointing capabilities to 
measure 
radiation from nebulas with high sensitivity, but the next-generation 
gamma-ray detectors (such as the Advanced Compton Telescopes) 
should be able to perform such measurements.

Another effect we have investigated is the overall contribution of 
PAHs to the 511 keV line from the ISM, recalling that we now have 
high-quality spectra of the annihilation emission from 
the galactic center region measured by INTEGRAL-SPI \citep{J06}.
We attempted to relate the reaction rate we have calculated to 
the annihilation spectra that have been reduced from the INTEGRAL-SPI 
data. We applied the same method as in \citet*{J06}, but instead of 
fitting the contribution of grains ($x_{gr}$ in Eq. 3 of 
\citealt{J06}) 
we fit the parameter $x_{PAH} = Y_{PAH} \langle f_{elec} \ \rangle$ 
to the 
measured spectrum. We obtained an upper-limit $x_{PAH} < $ 3.0 
$\times$ 10$^{-7}$. This upper-limit is in the range of expected 
values of $Y_{PAH}$ and $\langle f_{elec} \ 
\rangle$. However, the annihilation rate per unit density of 
PAH used for this fit is affected by an uncertainty of 47\%. 
Therefore, the formal upper-limit should be $x_{PAH} < $ 4.6 $\times$ 
10$^{-7}$. 
This defines a limit in the $Y_{PAH}$ -- $\langle f_{elec} 
\rangle$ space beyond which these parameters are not allowed, or else 
the effect would have been seen in the spectral analysis of the 
annihilation emission measured by SPI. Figure 4 shows the 
authorized domain $x_{PAH} = Y_{PAH} \times \langle f_{elec} 
\rangle <$ 4.6 $\times$ 10$^{-7}$. For instance, if all PAHs are 
negatively charged, then their abundance in the Galactic bulge should 
be 
less than 1.3 $\times$ 10$^{-7}$. When all PAHs are neutral, their 
abundance should be less than 4.6 $\times$ 10$^{-7}$. 

This result is very consistent with other estimations, which are 
obtained from totally different approaches: using analysis of 
Infrared Space Observatory (ISO) observations of the PAH emission in 
the Galaxy, \citet*{Wol03} obtain a total PAH abundance of 6 $\times$ 
10$^{-7}$; from their spectral analysis of [SiPAH]$^+$ complexes (the 
6.2 $\mu$m AIB), \citet*{Joa09} infer a PAH-to-H abundance of 8 
$\times$ 10$^{-7}$. We should note, however, that such measurements 
are not from the galactic bulge, where the presence and amounts of 
PAHs is unknown, so we cannot in all rigor compare the two results, 
despite their tantalizing closeness. Indeed, detections of PAH IR 
emission has been done from almost everywhere in the Galaxy (see 
earlier references plus \citealt{Gia88, Gia89}, and \citealt{Gia94}), 
but not the bulge itself ($l < 8 \deg)$; in the bulge, however, 
emission features of PAH molecules have been detected in planetary 
nebulae (\citealt{Per09, Phi09}).

%%%%%%%% Figure 4: Constraint on the Ypah-<f> domain %%%%%%%%
\begin{figure*}
\centering
\includegraphics[width=11.73cm,height=8.0cm]{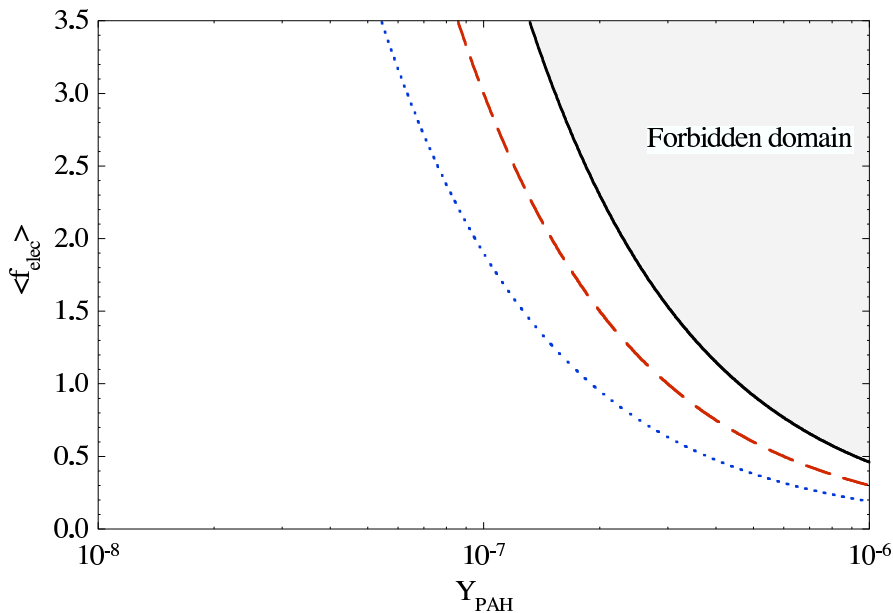}
\caption{Domain of $Y_{PAH}$ and $\langle f_{elec}\rangle$ values 
authorized by the SPI data: $x_{PAH} < $ 4.6 $\times$ 10$^{-7}$. 
The dashed and dotted lines correspond to $x_{PAH}$ = 3.0 
$\times$ 10$^{-7}$ and  $x_{PAH}$ = 1.9 $\times$ 10$^{-7}$, 
respectively.}
\end{figure*}
%%%%%%%%%%%%%%%%%%%%%%%%%%%%%%%%%%%%%%%%%%%

%%%%%%%%%%%%%%%%%%%%%%%%%%%%%%%%%%%%%%%%%%%%%%%%%%%%%%%%%%%%%%%%%%%%%%%% 
% 6. Summary and Conclusions
%%%%%%%%%%%%%%%%%%%%%%%%%%%%%%%%%%%%%%%%%%%%%%%%%%%%%%%%%%%%%%%%%%%%%%%% 

\section{Summary and Conclusions}

The contribution of PAHs to the annihilation of positrons in the ISM 
had not been considered up till now. The simple reason for this is 
that PAHs have always been considered as the smallest type of dust 
grains, and since the cross section for positron annihilation on 
grains had been shown \citep{Zur85} to be geometrical (modulo 
electric and 
destruction effects), the contribution of the (extremely small) PAHs 
was dismissed as negligible \citep{GJG05}. The error in that 
consideration was in overlooking the huge resonances in positron-PAH 
reactions (which, furthermore, did not occur for electrons), 
resonances that were vibrational in nature and led to increases in 
the annihilation cross sections by factors of up to 10$^7$. 
This is what prompted us to investigate this effect in its potential 
astrophysical applications.

Our current knowledge of e$^+$--PAH reactions is still very 
limited, both from experimental and theoretical perspectives. We have 
thus undertaken to estimate the rate of positron annihilation on PAH 
molecules in the ISM by a semi-empirical method. We have used all the 
information available to us from laboratory 
measurements of cross sections and from space IR analyses and 
estimates of abundances and distributions of PAHs in the ISM. We 
have had to make some extrapolations, which we have shown to be 
uncertain but
rather reasonable, from alkane data to PAH values and from low 
temperatures to ISM temperatures. 

We have found that large molecules (with a total number of atoms of 
about 50) are most important for positron annihilation. We have also 
obtained a total rate of positron annihilation on PAH molecules, and 
we have found that if the total amount of PAH (relative to hydrogen) 
in the ISM is of the order of 10$^{-6}$, then this process becomes 
second in importance in the warm neutral and the warm ionized phases, 
especially if the molecules are negatively charged. 
The issue of the charge states of the PAHs in various ISM conditions 
complicates the problem somewhat; indeed, depending on the 
temperature but particularly on the intensity of the UV field in the 
region, the PAHs may be neutral, positively, or negatively charged, 
and this would enhance or reduce their reaction rates with the 
positrons.

We have emphasized the large uncertainties inherent in this first 
attempt at tackling the problem, and we have pointed to the necessary 
future experimental, observational, and theoretical work for a 
substantial improvement of our understanding of this issue. In 
particular, we have called for experimental measurements of cross 
sections of positron annihilation on large PAH molecules. We also 
hope that future IR research, particularly with the Spitzer 
telescope, will help determine more precisely the specific PAHs which 
exist in space as well as their relative abundances in the ISM.

Finally, we have pointed to at least one future observational test, 
namely the measurement of the 511 keV line from specific nebulas 
where PAHs may be particularly abundant and where physical conditions 
may have led to substantial negative charging of the molecules and 
hence to enhanced positron annihilation rates. As we explained, such 
observational tests cannot be undertaken with current gamma-ray 
instruments, but the next-generation gamma-ray detectors should be 
able to. We have also attempted to relate the reaction rate we have 
calculated to the annihilation spectra that have been reduced from 
the INTEGRAL-SPI data. We have shown that the spectral analysis may 
provide constraints on the abundance and the electric charge of PAH 
in the ISM. Using the semi-empirical model of annihilation of 
positrons with PAH, 
the analysis of the spectrum of the annihilation emission from the 
galactic bulge measured by SPI tells us that the number-abundance 
of PAH relative to hydrogen should be less that 4.6 $\times$ 
10$^{-7}$ if all PAHs are neutral and even lower if many of them are 
negatively charged.

%%%%%%%%%%%%%%%%%%%%%%%%%%%%%%%%%%%%%%%%%%%%%%%%%%%%%%%%%%%%%%%%%%%%%%%%%
% Acknowledgements
%%%%%%%%%%%%%%%%%%%%%%%%%%%%%%%%%%%%%%%%%%%%%%%%%%%%%%%%%%%%%%%%%%%%%%%%%

\section*{Acknowledgments}

N. Guessoum acknowledges financial support from the American 
University of Sharjah through a research grant for this work and the 
Centre d'Etude Spatiale des Rayonnements for its hospitality and 
services during the conduct of this research. The authors thank Cliff 
Surko for pointing out the complications in estimating the rates at 
higher temperatures due to the change in the behavior of the resonant 
reactions above vibrational frequencies, Alain Kotz for reviewing and 
correcting our information on PAHs, Christine Joblin for useful 
discussions and pointers, and Paolo Pilleri for providing us with the 
PAH size distribution of Draine \& Lazarian (1998) in tabular form.

\end{document}